\documentclass[final,3p,times,twocolumn]{elsarticle}

\usepackage{graphicx}

\usepackage{amssymb}

\usepackage{natbib}

\journal{Physica B}

\begin{document}

\begin{frontmatter}



\title{Exploring the Charge-Ordering Transition in (TMTTF)$_2$X via Thermal Expansion Measurements}

\author[FFM]{Mariano de Souza}
\author[FFM]{Daniel Hofmann}
\author[Orsay]{Pascale Foury-Leylekian}
\author[Orsay]{Alec Moradpour}
\author[Orsay]{Jean-Paul Pouget}
\author[FFM]{Michael Lang}

\address[FFM]{Physikalisches Institut, Goethe-Universit\"{a}t
Frankfurt, SFB/TRR 49, D-60438 Frankfurt(M), Germany}
\address[Orsay]{Laboratoire de Physique des Solides, Universit\'{e} Paris Sud, CNRS UMR 8502, Orsay, France}

\begin{abstract}
We report results of high-resolution measurements of the
\emph{c$^*$}-axis expansivity ($\alpha_{c^{*}}$) at the
charge-ordering (CO) transition for the quasi-1D (TMTTF)$_{2}$X
compounds with X = SbF$_6$ and Br and make a comparison with
previous results for the X = PF$_6$ and AsF$_6$ salts. For X =
SbF$_6$, due to the screening of the long-range Coulomb forces,  a
sharp $\lambda$-type anomaly is observed at $T_{CO}$, which
contrasts with the step-like mean-field anomaly at $T_{CO}$ for
PF$_6$ and AsF$_6$, where CO occurs in the Mott-Hubbard charge-localized regime. For the latter two salts, a negative contribution
to $\alpha_{c^{*}}$ is observed above $T_{CO}$. This feature is
assigned to the anions' rigid-unit modes, which become inactive for
$T$ $<$ $T_{CO}$. Our $\alpha_{c^{*}}$ results for the X = Br salt, where such
rigid-unit modes are absent, reveal no traces of such negative
contribution, confirming the model based on the
anions' rigid-unit modes for the X = PF$_6$ and AsF$_6$ salts.

\end{abstract}


\end{frontmatter}


\section{Introduction}
\label{Introduction} The observation of a charge-ordering (CO)
transition \cite{Chow9a3} coinciding with the onset of
ferroelectricity \cite{Monceau2001-20} in the quasi-1D conductors of
the (TMTTF)$_2$X family with X = PF$_6$ ($T_{CO}$ $\simeq$ 65\,K
\cite{Chow9a3}), AsF$_6$ ($T_{CO}$ $\simeq$ 105\,K \cite{Chow9a3}),
SbF$_6$ ($T_{CO}$ $\simeq$ 154\,K \cite{Monceau2001-20}) salts have
revealed the exceptional properties of the Mott-Hubbard insulating
phase in these materials. Also for the X = Br salt, some experimental
evidence for a CO transition around 50\,K have been reported in the
literature \cite{Coulon2007}. The CO transition has been assigned to
the importance of both on-site $U$ and inter-site $V$ Coulomb
interactions \cite{Seo-1997}, with strong influence from
electron-lattice coupling
\cite{Riera2001,Clay-2003,Brazowskii-2003}. Actually, first
indications for a CO transition in the (TMTTF)$_2$X family were
observed more than twenty years ago. Resistivity measurements on the
X = SbF$_6$ salt \cite{Laversanne84} revealed a sharp
metal-to-insulator transition at $T_{MI}$ $\simeq$ 154\,K,
coinciding with a pronounced change in the thermopower
\cite{Coulon85} and a peak divergence in the dielectric permittivity
$\epsilon'$ \cite{Javadi88}. The absence of signatures in the magnetic
susceptibility \cite{Coulon85} led to the belief that only charge
degrees of freedom are involved in this transition. Furthermore, due
to the experimental difficulties posed by detecting lattice effects
associated with the CO transition \cite{Foury2002}, no evidence of
structural changes \cite{Laversanne84} has been reported and, due to
this, the CO transition has been labeled \emph{structureless}
transition. By means of ultra-high-resolution thermal expansion
experiments, we recently succeeded in detecting strongly anisotropic
lattice effects accompanying the CO transition for the X = PF$_6$
and AsF$_6$ salts \cite{Mariano08}. Most importantly, for the latter
salts, the strongest effects at both the charge ordering and the
spin-Peierls transition \cite{Mariano09} occur along the
\emph{c$^*$}-axis, along which planes of (TMTTF)$_{2}^{+}$ molecules
alternate with planes of counter anions X$^{-}$ \cite{Pouget1996}.
Hence, the results in \cite{Mariano08} provide the first
experimental evidence of the role of anion displacements for the
stabilization of the charge-ordered phase, in line with theoretical
predictions employing the extended Hubbard model coupled to the
anions \cite{Riera2001}. Here we report on thermal expansion
measurements along the \emph{c$^*$}-axis of (TMTTF)$_{2}$X with X =
SbF$_6$ and Br and make a comparison with previous results for X =
PF$_6$ and AsF$_6$.\label{intro}

\section{Experimental}\label{Experimental}
High-quality single crystals of (TMTTF)$_2$X with centrosymmetric
anions X = SbF$_6$ and Br were grown from THF using the standard
constant-current procedure. The uniaxial thermal expansion
coefficient, $\alpha(\textit{T})=\textit{l}^{-1}(\partial
\textit{l}/\partial \textit{T})$, $l$ denoting the sample length,
was measured by employing an ultra-high-resolution capacitance
dilatometer with a maximum resolution of $\Delta l/l=10^{-10}$,
built after \cite{Pott20a}. The dilatometer used here has proven
particularly useful for exploring lattice effects in organic
charge-transfer salts, see, e.g.\,\cite{de Souza 07}. The
experimental data presented were corrected for the thermal expansion
of the dilatometer cell with no further data processing. The
alignment of the crystal orientation was guaranteed with an error
margin of $\pm$5$^o$. \label{experimental}

\section{Results and Discussion}\label{Results}
\begin{figure}[!ht]
\begin{center}
\includegraphics[angle=0,width=0.48\textwidth]{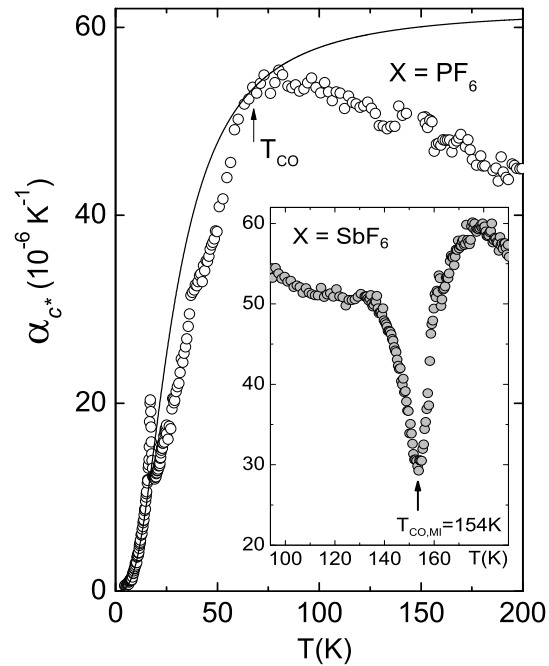}
\end{center}
\caption{Uniaxial expansitivity along the \emph{c$^*$}-axis of the
(TMTTF)$_2$X salt with X =  PF$_6$ (data taken from
Ref.\,\cite{Mariano08,Mariano09}). Solid line represents an
ordinary Debye-like behavior. Inset: Data for the X = SbF$_6$ salt
around the CO transition.}\label{Fig-1}
\end{figure}

In the main panel of Fig.\,\ref{Fig-1} we recall the results of
thermal expansion measurements along the \emph{c$^*$}-axis for the (TMTTF)$_2$X
salt with X = PF$_6$ \cite{Mariano08,Mariano09}. A similar behavior
is observed for the X = AsF$_6$ salt \cite{Mariano08,Mariano09} with
slight shifts in the positions of the anomalies at $T_{CO}$ and the
spin-Peierls transition $T_{SP}$. A striking feature revealed by
these experiments was the decrease of $\alpha_{c^{*}}$ for $T \geq
T_{CO}$, at variance with an ordinary lattice expansion which
monotonously increases with increasing temperatures in a more or
less Debye-like manner, indicating the action of a negative
contribution to $\alpha_{c^{*}}$ in this temperature regime. In
Ref.\,\cite{Mariano08}, this effect has been attributed to
rigid-unit modes of the PF$_{6}$ and AsF$_{6}$ anions and their
coupling to the charge ordering: on approaching $T_{CO}$ from above,
growing CO fluctuations, cause, via S-F contacts, positional
fluctuations of the anions towards their new off-center equilibrium
positions. These positional fluctuations provide an effective
damping of the anions' rigid-unit modes. Upon cooling through
$T_{CO}$, the CO becomes static, giving rise to a freezing of these
vibration modes and, as a consequence, the negative contribution in
$\alpha_{c^*}$ is no longer active.

\begin{figure}[!ht]
\begin{center}
\includegraphics[angle=0,width=0.48\textwidth]{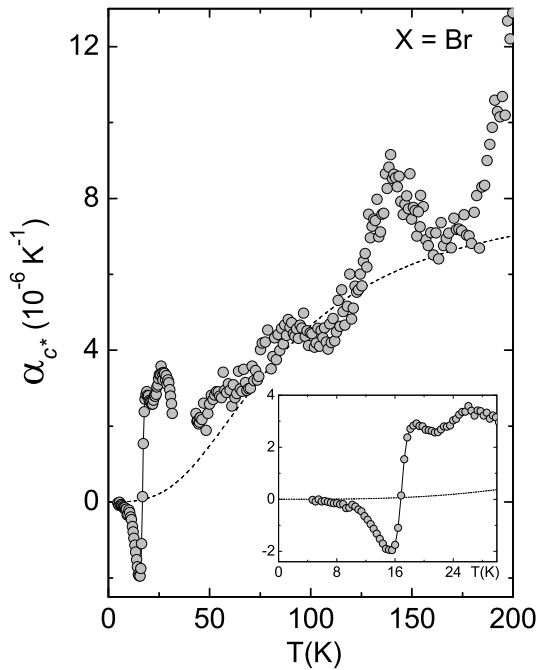}
\end{center}
\caption{Uniaxial expansitivity along the \emph{c$^*$}-axis of the
(TMTTF)$_2$X salt with X =  Br. Dashed line represents an ordinary
Debye-like behavior. Gap between 32 and 42\,K corresponds to a range
of enhanced noise.  Inset: Blowup of the low-temperature
data.}\label{Fig-2}
\end{figure}

In Fig.\,\ref{Fig-2} we show corresponding data for the X = Br salt,
where no such rigid-unit modes exist. In fact, for the temperature
range shown here, the data lack any indication for a negative
contribution in $\alpha_{c^*}$, consistent with the model proposed in
\cite{Mariano08}. Instead, $\alpha_{c^*}$ shows distinct anomalies,
the origin of which are not fully understood at present. This refers
to the peak anomaly centered around 140\,K, reproduced for the two
crystals investigated here, as well as the upturn above $\thicksim$
180\,K. Another remarkable aspect is that $\alpha_{c^*}$ is roughly
one order of magnitude smaller in the X = Br than in the PF$_6$ and
AsF$_6$ salts. A possible explanation for this discrepancy as well
as for the above-mentioned absence of the negative contribution to
$\alpha_{c^*}$ for the X = Br salt could be associated to the more dense
packing of the X = Br salt along the ${c^*}$-axis. In other words,
the Br$^-$ anion fits better into the methyl group cavities
\cite{Pouget1996}. This leaves less possibility to a CO transition
in this salt triggered by anion displacements, as it occurs in X =
PF$_6$, AsF$_6$ \cite{Mariano08}.

For the X = SbF$_6$ salt (inset of Fig.\,\ref{Fig-1}), a
$\lambda$-like negative anomaly is observed at $T_{CO}$ = 154\,K
which nicely coincides with the peak anomaly in $\epsilon'$
\cite{Nad2006-20}. The shape and size of the phase transition
anomaly, indicative of the presence of strong fluctuations, differ
markedly from the mean-field-like anomaly at $T_{CO}$ for the X =
PF$_6$, AsF$_6$ salts, cf.\,Fig.\,\ref{Fig-1} and
Ref.\,\cite{Mariano08}. This difference can be understood as a
consequence of short-range Coulomb forces in the SbF$_6$ salt, where
CO coincides with a metal-insulator transition ($T_{CO}$ =
$T_{MI}$), implying a more effective screening of the long-range
Coulomb forces above $T_{CO}$, as compared to the AsF$_6$ and PF$_6$
salts, where $T_{CO} < T_{\rho}$, with $T_{\rho}$ denoting the
position of the minimum in the resistivity, which marks the onset of
the charge localization \cite{Dressel07}.

The negative sign of the phase transition anomaly indicates,
according to the Ehrenfest relation, a decrease of $T_{CO}$ upon
applying uniaxial pressure along the $c^{*}$ axis,
i.e.\,$dT_{CO}$/$dP_{c^*}$ $<$ 0. In fact, $^{13}$C-NMR measurements
on the X = SbF$_6$ under hydrostatic pressure revealed a negative
pressure dependence of $T_{CO}$ \cite{Zamborsky2002-20}. Hence, the
present results suggest that also for the X = SbF$_6$ salt the
\emph{c$^*$} lattice parameter plays an important role in the
pressure-induced changes of $T_{CO}$. The results for the X =
SbF$_6$ salt in the whole temperature range will be published
elsewhere \cite{MdeSouza2009}.

We turn now to the anomalous behavior of the X = Br salt at low
temperatures, see main panel and inset of Fig.\,\ref{Fig-2}. The crucial observation made here is a dramatic deviation from
a Debye-like behavior in $\alpha_{c^*}$ below $\sim$\,50\,K,  accompanied by a pronounced step-like anomaly around 17\,K  and
a change of sign. Remarkably, according to Ref.\,\cite{Parkin-1982},
the transition into the antiferromagnetically ordered ground state
of this salt takes place at 13\,K, i.e.\,somewhat below the
discontinuity in $\alpha_{c^*}$. We stress that below 17\,K, a rapid
drop of the 2\emph{k}$_F$ bond fluctuations has been reported in
\cite{Pouget-1997} as well as a plateau in the resistivity
centered around the same temperature \cite{Dressel07}. In addition,
below $T$ $\simeq$ 20\,K the conductance becomes frequency dependent
\cite{Nad-98}, signalizing a relaxor-type of behavior. Since  a
broad maximum in $\epsilon'$ is observed in the 35-50\,K temperature
range \cite{Nad-98}, the possibility of the formation of a local CO
and/or ferroelectric domains cannot be ruled out. Further
experiments are needed to better understand the anomalous
low-temperature regime of this salt.

In conclusion, by measurements of the uniaxial expansivity along the
$c^*$-axis of various members of the (TMTTF)$_2$X family, distinct
lattice effects accompanying the charge-ordering transition have
been observed. From the different character of the phase transition
anomaly for the various salts we infer the presence of short-range
Coulomb forces for X = SbF$_6$, which contrasts with the
longer-ranged Coulomb interactions, i.e. a reduced screening, for X
= PF$_{6}$ and AsF$_{6}$. The X = Br compound  lacks any indications
for a negative contribution to the $c^*$-axis expansivity, which for
the PF$_{6}$ and AsF$_{6}$ salts have been attributed to positional
fluctuations  of these anions \cite{Mariano08}.



\end{document}